\documentclass[10pt,pra,aps,twocolumn,showpacs,superscriptaddress]{revtex4-1}
\usepackage{amsmath}
\usepackage{latexsym}
\usepackage{amssymb}
\usepackage{graphics,epstopdf}
\usepackage{graphicx}
\usepackage[colorlinks=true,citecolor=blue, urlcolor=blue]{hyperref}
\usepackage{float}

\begin{document}

\title{Local Orthogonality provides better upper bound for Hardy's nonlocality}

\author{Subhadipa Das}
\email{sbhdpa.das@bose.res.in}
\affiliation{S.N. Bose National Center for Basic Sciences, Block JD, Sector III, Salt Lake, Kolkata-700098, India}

\author{Manik Banik}
\email{manik11ju@gmail.com}
\affiliation{Physics and Applied Mathematics Unit, Indian Statistical Institute, 203 B.T. Road, Kolkata-700108, India}

\author{Md. Rajjak Gazi}
\email{rajjakgazimath@gmail.com}
\affiliation{Physics and Applied Mathematics Unit, Indian Statistical Institute, 203 B.T. Road, Kolkata-700108, India}

\author{Ashutosh Rai}
\email{arai.qis@gmail.com}
\affiliation{CAPSS, Department of Physics, Bose Institute, Sector-V, Salt Lake, Kolkata 700 091, India}

\author{Samir Kunkri}
\email{skunkri@yahoo.com}
\affiliation{Mahadevananda Mahavidyalaya, Monirampore, Barrakpore, North 24 Parganas-700120, India}

\begin{abstract}
The amount of nonlocality in quantum theory is limited compared to that allowed in generalized no-signaling theory [\href{http://link.springer.com/article/10.1007\%2FBF02058098}{Found. Phys. {\bf24}, 379 (1994)}]. This feature, for example, gets manifested in the amount of Bell inequality violation as well as in the degree of success probability of Hardy's (Cabello's) nonlocality argument. Physical principles like information causality and macroscopic locality have been proposed for analyzing restricted nonlocality in quantum mechanics---\emph{viz.} explaining the Cirel'son bound. However, these principles are not that much successful in explaining the maximum success probability of Hardy's as well as Cabello's argument in quantum theory. Here we show that, a newly proposed physical principle namely Local Orthogonality does better by providing a tighter upper bound on the success probability for Hardy's nonlocality. This bound is relatively closer to the corresponding quantum value compared to the bounds achieved from other principles.         
\end{abstract} 

\maketitle
\section{Introduction}
All the correlations between outcomes of measurements performed on spatially separated parts of a composite quantum system cannot be described in a {\it local realistic} framework and hence are called nonlocal \cite{EPR, Bell}. Such correlation in quantum mechanics (QM) is witnessed by violation of some conditions like the celebrated  Bell-Clauser-Horne-Shimony-Holt (B-CHSH) inequality\cite{Bell,CHSH}, Hardy's nonlocality argument \cite{Hardy1} and Cabello's nonlocality argument \cite{Cabello1,Kunkri1,Lin} which are derived under the assumption of \emph{local realism}. It is known that, though QM violates these conditions, it satisfies the No-Signaling (NS) principle. Popescu and Rohrlich first showed that there exist correlations (e.g. PR-box correlation) which are more nonlocal than quantum correlations but still satisfy the NS principle \cite{Popescu}. While in a generalized non-signaling theory (GNST) \cite{Hardy3} B-CHSH expression can reach the maximum algebraic value $4$, in QM this value is restricted by the Cirel'son bound ($2\sqrt{2}$) \cite{Cirel'son}, expressing a limit on the amount of nonlocality within quantum mechanics. Such limits on quantum nonlocality is also observed in the maximum success probability of Hardy's nonlocality argument as well as in the maximum success probability of Cabello's nonlocality argument (henceforth abbreviated as HNA and CNA respectively). It has been shown that in GNST maximum success probability for both HNA and CNA is $0.50$ \cite{Kunkri2, Cereceda}, whereas in QM the corresponding values are $\approx0.09$ \cite{Hardy1,Hardy2, Rabelo} and $\approx0.11$  \cite{Kunkri1,foot1}.  

In recent years, several attempts have been taken to explain the restricted nonlocal features in QM from some physical or information theoretical principles. Initially, Van Dam showed that distant parties having access to the PR-box correlation can render communication complexity trivial and then argued that this could be a reason for the non-existence of such stronger nonlocal correlations in nature \cite{Van Dam}; another progress along this line was reported in \cite{Brassard}. Further important breakthroughs were obtained by introducing physical principles like Information Causality (IC) \cite{Pawlowski} and Macroscopic Locality (ML) \cite{Navascues1} which explained the Cirel'son bound in QM. It has also been shown that this bound can  also be explained by the uncertainty principle (along with steering) \cite{Oppenheim} as well as by complementarity principle \cite{Manik}.  

Though considerable amount of successful attempts have been made in explaining the restricted violation of B-CHSH inequality in QM, there has been little progress in explaining the restricted success probabilities of HNA and CNA in QM. In \cite{ali}, it has been shown that a necessary condition for respecting the IC principle restricts maximum success probability for both HNA and CNA to $0.207$. However, this value is related with the Cirel'son bound as it corresponds to the maximum success probability over set of NS correlations for which the B-CHSH violation is restricted to $2\sqrt{2}$ \cite{xiang}. Thus, the known IC condition though lowers the bound on success probabilities for HNA and CNA compared to the bound imposed by the NS principle only, this bound still remains related to optimal Bell violation in QM. On the other hand, as we have checked, under the condition of ML, the maximum success probability of the HNA (CNA) remain nearly same (slightly lower) as that achieved in \cite{ali}. 

More recently, it has been shown that no physical principle which is bipartite in nature, for example IC and ML, can completely identify the full set of quantum correlations \cite{Gallego,Das}. For this purpose, a newly proposed physical principle namely \emph{Local Orthogonality} (LO) \cite{Fritz1} (also known as \emph{Global Exclusivity}\cite{Cabello2}) has drawn much interest \cite{Fritz1,Cabello2,Cabello3,Yan}. The principle of LO states that events involving different outcomes of the same local measurement must be orthogonal (or exclusive) and that the sum of probabilities of pairwise orthogonal events cannot exceed $1$. This principle has an interesting feature: it may happen that single copy of a given correlation satisfies LO whereas two or more copies of this correlation violates LO. Although, unlike IC or ML, the LO principle still seeks to single out the Cirel'son bound; it has been shown that LO when applied with two copies of correlations can produce tighter bound on nonlocality (B-CHSH violation) and goes closer to the Cirel'son bound \cite{Fritz1,Cabello2}. So there remains a scope to derive exact Cirel'son bound by considering LO with multiple copies of correlations. Moreover, LO is a genuine multipartite principle with greater scope for characterizing multipartite nonlocality. For example, LO principle can exclude all extremal no-signaling nonlocal correlations in tripartite case. In this paper, we show the strength of LO principle by considering two copies of a Hardy's correlation to find that maximum success probability of HNA is significantly lower compared to those derived under consideration of IC and ML. In this respect, however, the obtained bound for Cabello's nonlocality remains same as that derived from the IC principle. 

The paper is organized in the following way. In section-\ref{ns} we describe general no-signaling framework for a bipartite two-level system and give the formulation of HNA and CNA. In section-\ref{ml} we find the maximum success probability of HNA and CNA under the principle of Macroscopic Locality. In section-\ref{lo} we first briefly review the LO principle and then find the maximum success probability of HNA and CNA under consideration of LO constraints on two copies of the correlation. Section-\ref{conclusion} contains brief discussion and conclusions of our work.


\section{No-Signaling Framework and Hardy-Cabello Nonlocality }\label{ns}
For two spatially separated observers (say Alice and Bob), let $P(ab|xy)$ denotes probability of obtaining outcome $a$ at Alice's end and outcome $b$ at Bob's end conditioned that measurement $x$ and $y$ are performed by Alice and Bob respectively $(x,y,a,b \in\{0,1\})$. The set of bipartite binary input-output probability distribution form a $8$ dimensional polytope \cite{Khalfin,Pironio} and such a distribution can be expressed in terms of $8$ parameters (see Table-\ref{table1}). 
\begin{center}
\begin{table}[h!]\label{table1}
\begin{tabular}{|c|c|c|c|c|}
	\hline
$xy\backslash ab$ &  $0_A0_B$    &   $0_A1_B$  & $1_A0_B$   & $1_A1_B$  \\
	\hline\hline
$0_A0_B$  & $c_1$& $m_0-c_1$& $n_0-c_1$ & $1+c_1-m_0-n_0$ \\
\hline
$0_A1_B$   & $c_2$& $m_0-c_2$& $n_1-c_2$ & $1+c_2-m_0-n_1$\\
\hline
$1_A0_B$   & $c_3$& $m_1-c_3$& $n_0-c_3$ & $1+c_3-m_1-n_0$\\
\hline
$1_A1_B$  & $c_4$& $m_1-c_4$& $n_1-c_4$ & $1+c_4-m_1-n_1$\\
\hline
\end{tabular}
\caption{Bipartite two input-two output NS probability distribution. For $i\in \{0,1\}$, $m_i~(n_i)$ is the Alice's (Bob's) marginal probability of obtaining outcome `$0$' for measurement `$i$'. Due to positivity we have $\mbox{max}\{0,m_0+n_0-1\}\le c_1\le \mbox{min}\{m_0,n_0\}$; and similar conditions holds for other $c_j$, $j\in \{2,3,4\}$.}\label{table1}
\end{table}
\end{center}
\textbf{Hardy-Cabello Nonlocality Argument}---Now, let us consider joint probabilities satisfying the following constraints:
\begin{eqnarray}\label{h1}
P (01|xy) &=& q_1, \\\label{h2}
P (00|\bar{x}y) &=& 0, \\\label{h3}
P (10|x\bar{y}) &=& 0, \\\label{h4}
P (00|\bar{x}\bar{y}) &=& q_4 ,
\end{eqnarray}
where $\bar{\alpha}$ denotes complement of $\alpha$. These equations form the basis of Hardy-Cabello nonlocality argument. It can easily be seen that these equations contradict local realism if $q_1 < q_4$ . To show this, let us consider those hidden variable states $\lambda$ for which both the  measurement $\bar{x}$ and $\bar{y}$ achieve outcome `$0$'. For these states, Eq.$(\ref{h2})$ and Eq.$(\ref{h3})$ tell that the outcomes of $x$ and $y$ must be equal to `$0$' and `$1$', respectively. Thus according to local realism $ P(01|xy)$ should be at least equal to $q_4$. This contradicts Eq.$(\ref{h1})$ as $q_1 < q_4$. Whenever $q_1 = 0$, the Cabello's argument reduces to Hardy's argument. So by Cabello's argument, we specifically mean that the above argument runs, even with nonzero $q_1$. One important difference between HNA and CNA is that a mixed two-qubit entangled state can never exhibit HNA \cite{Kar}, but they can exhibit CNA. 

It is interesting to note that while in QM maximum success probability of HNA and CNA is $0.09$ \cite{Hardy1,Hardy2,Rabelo} and $0.11$ \cite{Kunkri1} respectively, in GNST there exist correlation that exhibits HNA and CNA with maximum success probability $0.5$. Thus, like B-CHSH violation, Hardy's as well as Cabello's maximum success probabilities in QM are also limited in comparison to GNST. Interestingly, in \cite{ali} the authors observed that under the necessary condition for satisfying IC principle the maximum success probabilities for HNA as well as for CNA reduce to $0.207$.

In the remaining part of this paper, without loss of generality we consider the following form of Hardy-Cabello correlation:
\begin{equation}\label{hna}
P (01|01) = q_1,
P (00|11) = 0,
P (10|00) = 0,
P (00|10) = q_4 
\end{equation}
%
%
\section{Maximum Success of HNA \& CNA under ML}\label{ml}
The principle of Macroscopic Locality (ML) \cite{Navascues1} is known to be respected by quantum correlations. This principle can successfully reproduce the Cirel'son bound ($2\sqrt{2}$). The ML principle can be expressed as follows: consider a coarse grained Bell-type experiment, where a collection of $N$ sources $S_1, S_2,...,S_N$ emit one particle each in different branches and each branch is composed of a beam of $N$ particles. However, in any beam, the information about the specific source of any single particle is lost. Local measurement (same for each particle) records the average number of particles with different possible outcomes. The principle of Macroscopic Locality states that in the limit of large $N$, statistics generated from any coarse grained Bell-type experiment with the help of physically realizable sources must be local, i.e., it must satisfy any Bell-type inequality\cite{Navascues1}. In the bipartite case with two dichotomic observables at each site the necessary and sufficient criteria for respecting Macroscopic Locality become \cite{Navascues1,Navascues2,Navascues3}:
\begin{equation}\label{ml1}
|\sum_{x,y=0}^{1}(-1)^{xy}\sin^{-1}(D_{xy})|\leq \pi
\end{equation}
where $D_{xy}=\frac{\langle a_{x}b_{y} \rangle-\langle a_{x} \rangle.\langle b_{y}\rangle}{\sqrt{(1-\langle a_{x}\rangle^2)(1-\langle b_{y}\rangle ^2)}}$. For a general NS correlation we obtained from TABLE-\ref{table1} that:
\begin{eqnarray}\label{mlc}
\langle a_{0}b_{i}\rangle=1+4c_{1+i}-2(m_0+n_{0+i}),\\
\langle a_{1}b_{i}\rangle=1+4c_{3+i}-2(m_1+n_{0+i}),\nonumber\\
\langle a_{i}\rangle=2m_{0+i}-1,\langle b_i\rangle=2n_{0+i}-1,~~i\in\{0,1\}\nonumber.
\end{eqnarray}
Now, consider the Hardy-Cabello argument in Eq. (\ref{hna}), for $q_1=0$, we numerically maximize $q_4$ under Macroscopic Locality condition (given in Eq. (\ref{mlc})), the maximum success probability of HNA turns out to be $\approx0.2062$. Similarly, for non zero $q_1$, on maximizing $q_4-q_1$, the maximum success probability of CNA is $\approx0.2062$ which is same as that for HNA.

It turns out that for a general NS-correlation maximum probabilities of Hardy's success and Cabello's success achieved under consideration of necessary condition for respecting IC principle is nearly same to those obtained by applying ML principle (for ML bounds are slightly improved). Interestingly, in the following section we show that for explaining the restricted Hardy's nonlocality in QM, application of LO principle with two copies of Hardy's correlation gives much tighter bound compared to those obtained from IC and ML.
%
%
\section{Maximum SUCCESS OF HNA \& CNA UNDER LO PRINCIPLE}\label{lo}
As pointed out in \cite{Fritz1} the \emph{ Local Orthogonality} (LO) principle states that the sum of the probabilities of pairwise orthogonal (exclusive) events cannot exceed $1$ (this follows from Specker's observation \cite{Specker} and Boole's axiom \cite{Boole}). To express the LO principle explicitly, suppose the joint conditional probability distribution $P (a_1 . . . a_n |x_1 . . . x_n )$ representing the probability of obtaining outcomes $a_1 , . . . , a_n$ when measurements $x_1 , . . . , x_n$ are performed by $n$ respective parties. Two different events $e_1 = (a_1 . . . a_n |x_1 . . . x_n )$ and $e_2
  = (a'_1 . . . a'_n |x'_1 . . . x'_n )$ are orthogonal if for some $i$, $a_i \neq a'_i$ when $x_i = x'_i$. Then, according to the LO principle  
\begin{equation}\label{lo-eqa}
\sum_k P(e_k)\leq 1,
\end{equation}
Where the set of events $\{e_k\}$ are pairwise orthogonal and called orthogonal set. This can be represented as a graph, where an event correspond to a vertex and a pair of orthogonal vertex define an edge. A set of orthogonal vertex form a clique (a complete subgraph). A clique is maximal if it cannot be extended to another clique by including a new vertex. Clearly, any clique in the orthogonality graph of events gives rise to an LO inequality (like Eq.-$(\ref{lo-eqa})$). LO principle is a multi-party generalization of the NS principle and various interesting results appear in recent time on considering this principle \cite{Fritz1,Cabello2,Cabello3,Yan}. It detects all known extremal non-local boxes as supra-quantum, including those for which any bipartite principle fails \cite{Fritz1}. Set of correlations that satisfy this principle is very close to the set of quantum correlations. Although, the LO principle still does not single out the Cirel'son bound of the B-CHSH inequality \cite {Fritz1,Cabello2} but this principle singles out quantum contextuality \cite{Cabello2}. Now in what follows we find the maximum success probability of HNA and CNA under the LO principle.

\subsection{Maximum success of HNA}
\begin{figure}
\centering
\includegraphics[height=6cm,width=6cm]{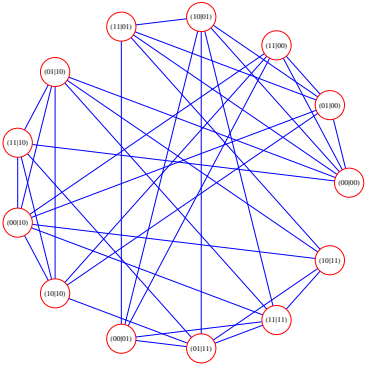}
\caption{This is the orthogonality graph corresponding to Hardy's correlation for a bipartite binary input-output scenario. In bipartite binary input-output scenario the events are of the form $(ab|xy)$ with $a,b,x,y\in\{0,1\}$. For Hardy's correlation among these $16$ events $3$ will never occur and thus orthogonality graph corresponding to the Hardy's correlation contains $13$ vertices. The edges connect two orthogonal events.}\label{fig1}
\end{figure}

For a bipartite binary input-output scenario the orthogonality graph corresponding to the Hardy's correlation contains $13$ vertices as shown in Fig.(\ref{fig1}). It is known that, in a bipartite scenario LO constrains on one copy of correlation is equivalent to NS conditions \cite{Fritz1}. So here, we consider two copies of Hardy's correlation and apply the LO principle. The conditional probability distributions for two-copy correlation is given by: 
\begin{equation}\label{product}
P(a_1b_1a_2b_2|x_1y_1x_2y_2)=P(a_1b_1|x_1y_1)P(a_2b_2|x_2y_2)
\end{equation}
The orthogonality graph (achieved by co-normal product of $2$ copies of orthogonality graph) now contains $169$ vertices. First, by using the software packages \cite{soft1,soft2}, we find all the maximal cliques for this graph. Then, for obtaining the maximum success of Hardy's argument under the full set of resulting LO inequalities, we find by writing a C program that it is sufficient to maximize Hardy's success probability under a small subset of LO inequalities (\ref{A11}) to (\ref{A20}) corresponding to the set of maximal cliques (\ref{A1}) to (\ref{A10}) given in the Appendix-\ref{apend1}. Therefore, now the problem reduces to:
\begin{eqnarray}
\mbox{Maximize}~ q_4 ~(=c_3) \nonumber \\
\mbox{subject~to~the~LO~inequalities~} (\ref{A11}) \mbox{~to~} (\ref{A20});  \nonumber\\
 0\leq p(ab|xy)\leq 1,\forall x,y,a,b;\nonumber\\
\mbox{~and,~} c_4=m_0-c_2=n_0-c_1=0.
\end{eqnarray}

Using \emph{MATHEMATICA}, we find that maximum success of Hardy's nonlocality argument turns out to be $0.177$. Clearly the obtained value is less than $0.207$ and relatively close to corresponding quantum value.

\subsection{Maximum success of CNA}
\begin{figure}
\centering
\includegraphics[height=6cm,width=6cm]{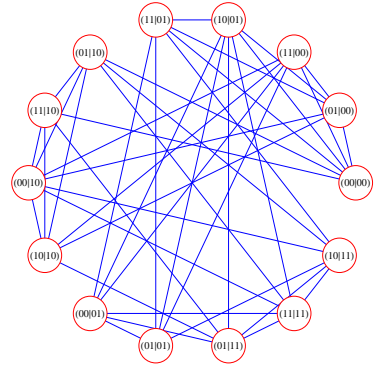}
\caption{This is the orthogonality graph corresponding to Cabello's correlation for a bipartite binary input-output scenario. For Cabello's correlation among these $16$ events $2$ will never occur and thus orthogonality graph corresponding to a general correlation contains $14$ vertices.}\label{fig2}
\end{figure}
For a bipartite binary input-output scenario the orthogonality graph corresponding to Cabello's correlation contains $14$ vertices as shown in Fig.(\ref{fig2}). With two copies, for Cabello's correlation there are $196$ vertices. In a similar way, for obtaining Cabello's maximum success probability under all LO inequalities, we find that it is sufficient to maximize under a subset of LO inequalities (\ref{B9}) to (\ref{B16}) corresponding to the set of cliques (\ref{B1}) to (\ref{B8}) given in the Appendix-\ref{apend2}. The problem thus reduces to
\begin{eqnarray}
\mbox{Maximize~} q_4-q_1~ (= c_3-m_0+c_2)\nonumber\\
\mbox{subject~to~the~LO~inequalities~} (\ref{B9}) \mbox{~to~} (\ref{B16});  \nonumber\\
0\leq p(ab|xy)\leq 1,\forall x,y,a,b;\nonumber\\
\mbox{~and,~} c_4=n_0-c_1=0.
\end{eqnarray}
After optimization, we obtain the maximum success for Cabello's argument $0.207$, which is same as the bound obtained from the IC Principle.
%
%
\section{CONCLUSIONS}\label{conclusion}
Explaining the restricted feature of Hardy's nonlocality in QM from some physical principle is an interesting open problem for a long time. In this respect, we first show that under ML principle the success probability for both Hardy's and Cabello's nonlocality reduces the success probability $0.5$ (no-signaling bound) to $0.206$ which is slightly improved from the bound $0.207$ derived from IC principle. Thus, bipartite principles like IC and ML, confine the maximum success probability of Hardy's (as well as Cabello's) argument in comparison to the maximum value allowed for a general no signaling distribution, but the value is still large compared to that in quantum theory. Further, we show that a very recently proposed physical principle called \emph{Local Orthogonality} gives better bound for Hardy's success probability compared to the bounds derived from IC or ML. Considering $2$-copy of a Hardy correlation and applying the LO constraints, we find that the maximum success probability of Hardy's arguments is $\approx 0.177$ which is relatively closer to the quantum value $\approx 0.09$. On the other hand, for Cabello's nonlocality similar consideration of the LO reduces the success probability $0.5$ (no-signaling bound) to $0.207$, which is same as that obtained from the IC principle. 

In quantum mechanics there is a genuine gap between the maximum success probability of Hardy's and Cabello's nonlocality argument but principles like NS, IC and ML fail to reveal this quantum feature. Interestingly, for a general no signaling correlation with two copies application of LO not only lowers the maximum Hardy's success towards quantum value, but it also reveals a gap between bounds on Hardy's and Cabello's success. Whether many-copy (possibly infinite) LO constrains can reduce success probability of Hardy's (Cabello's) argument exactly to the quantum value is an interesting open question. But answering this question is not easy as the problem of finding all the maximal cliques (which give LO inequalities in our case) of a graph are NP-hard. Our work does not solve the open problem concerning the restricted Hardy's nonlocality of QM, but it shows that LO  may be a potential candidate to explain this feature and it requires further research.

\begin{acknowledgments}
It is a pleasure to thank Guruprasad Kar for many stimulating discussion. SD and AR acknowledges support from the DST project SR/S2/PU-16/2007. AR also acknowledges the support from Bose Institute, Kolkata. 
\end{acknowledgments}
\begin{widetext}
\appendix
\section{A subset of maximal cliques for Hardy correlation}\label{apend1}
Following maximal cliques are sufficient for obtaining the maximum value of Hardy's success under the LO principle
\begin{eqnarray}
\left\{(0000|1010),(0101|1010),(0001|1010),(0100|1010),(0010|0011),(1100|1000),(1011|1101),(1111|0110),\right. \nonumber\\
\left.(1101|1110),(0110|1010)\right\} \label{A1}\\
\left\{(1010|0101),(1111|0001),(1100|1001),(0110|1111),(0011|0100),(1001|1000),(0000|0010)\right\} \label{A2} \\
\left\{(0101|1111),(0111|1111),(1110|0011),(0011|0100),(1011|0111),(1111|1111),(0000|1001),(1101|1110),\right. \nonumber \\ \left.(1100|1110),(1000|1110)\right\}\label{A3} \\
\left\{(0000|0101),(0010|0101),(1010|0101),(1000|0101),(1011|0111),(0011|0001),(1101|1011),(0100|1100),\right. \nonumber \\ \left.(1111|0110)\right\}\label{A4} \\
\left\{(0001|0111),(1100|0010),(0000|1001),(1101|1110),(0111|1100),(0110|1010),(1111|1011),(1110|1011),\right. \nonumber \\ \left.(1010|1011),(1011|1011)\right\}\label{A5} \\
\left\{(0001|0011),(0111|0001),(1011|1011),(0011|1011),(1111|0000),(0110|0001),(0101|1000),(1100|0110),\right. \nonumber \\ \left.(1000|1101),(0010|1011)\right\} \label{A6} \\
\left\{(0011|0001),(1010|1111),(1111|0110),(1011|0111),(0100|1100),(1100|1010),(1101|1010),
(0101|1010)\right\} \label{A7} \\
\left\{(0000|1010),(0111|1100),(1100|0001),(0011|0111),(1011|0111),(1111|1111),(0010|0111),(1110|1101),\right. \nonumber \\ \left.(1001|1111),(1101|1111)\right\}\label{A8}\\
\left\{(0011|0001),(0010|1011),(1100|0110),(1001|1000),(1000|1100),(0110|1010),(1111|1010),(0111|1010),\right. \nonumber \\ \left.(0101|1010),(1101|1010)\right\}\label{A9}\\ 
\left\{(0000|0001),(0011|0100),(0101|1111),(1100|1000),
(1111|0010),(1001|0110),(0110|1010)\right\}\label{A10}
\end{eqnarray}
Using Eq.(\ref{product}), the resulting LO inequalities can be expressed in terms of variables $c_k,m_i,n_i$ where $k\in\{1,2,3,4\}, i\in \{0,1\}$ corresponding to the above set of maximal cliques respectively. 
\begin{eqnarray}
c_3^2+2c_1n_1\leq c_1^2+n_1^2 \label{A11}\\
 2c_1c_3 \leq c_1^2+(c_2-n_1)(-1+m_1+n_1) \label{A12}\\ 
 c_2(c_3+m_1)+(c_3-m_1)n_1 \leq c_2^2  \label{A13}\\ 
c_3(1+m_1)+c_2(m_1+n_1)\leq c_2+m_1^2+c_3n_1 \label{A14}\\
c_1(-c_3+m_1)+c_3(c_3 + n_1)\leq m_1 n_1 \label{A15}\\
c_3 + c_1 c_3+c_2(-1+c_1-c_3+m_1+n_1)\leq c_1(m_1+n_1) \label{A16} \\ 
c_3+c_3^2+c_2(m_1+n_1) \leq c_2+c_3(m_1+n_1)\label{A17}\\
c_3^2+2c_2n_1 \leq c_2^2+n_1^2 \label{A18}\\ 
c_3(1+c_3)+c_1(c_2+m_1) \leq c_1+c_3(c_2+m_1)\label{A19}\\
c_3^2+m_1(-1+m_1+n_1) \leq (c_1-c_2)^2+c_3(-1+m_1+n_1)\label{A20}
\end{eqnarray}
\section{A subset of maximal cliques for Cabello correlation}\label{apend2}
Following maximal cliques are sufficient for obtaining the maximum value of Cabello's success under the LO principle
\begin{eqnarray}\label{ccli1}
\left\{(1010|0101),(0010|0101),(0110|0101),(1111|0110),(1101|0011),(0000|1000),(0011|0001),(0111|0001),\right. \nonumber \\ \left.(1000|1101),(0101|0001)\right\}\label{B1}\\
\left\{(1010|0101),(0010|0101),(0110|0101),(1111|0110),(0101|1111),(1100|1000),(1001|0100),(1011|0101),\right. \nonumber \\ \left.(0011|0101),(0000|0001)\right\}\label{B2}\\
\left\{(0000|1010),(0111|1110),(0110|1110),(0101|0110),(1101|0011),(1000|1101),(1111|1110),(1110|1110),\right. \nonumber \\ \left.(1011|0111),(0011|0100)\right\}\label{B3}\\
\left\{(1010|0101),(0010|0101),(0111|1110),(1100|0000),(1110|1101),(0001|1011),(0000|0101),(1111|1001),\right. \nonumber \\ \left.(1011|1011),(1001|1011)\right\}\label{B4}\\
\left\{(1010|0101),(0010|0101),(0110|0101),(1111|0110),(0001|1011),(1100|0000),(0000|0001),(0100|0001),\right. \nonumber \\ \left.(1011|1101),(0101|0001)\right\}\label{B5}\\
\left\{(0000|1010),(0011|0010),(0010|0010),(1001|1111),(1100|1001),(0110|0001),(0101|1000),(0100|1000),\right. \nonumber \\ \left.(1111|0100)\right\}\label{B6}\\
\left\{(0000|1010),(0011|0100),(0110|1111),(1111|0001),(0101|1001),(1110|1101),(1101|1001),(1100|1001),\right. \nonumber \\ \left.(1001|1001),(1000|1001)\right\}\label{B7}\\
\left\{((1010|0101),(0011|0110),(0001|0011),(1100|1000),(1111|0001),(1001|0100),(0100|1101),(0110|1101),\right. \nonumber \\ \left.(1110|1101)\right\}\label{B8}
\end{eqnarray}
Similarly, the corresponding LO inequalities constructed from above set of maximal cliques respectively are
\begin{eqnarray}
(1+c_1)c_3+c_2(1+c_2+c_3)+n_1(c_1+m_1+n_1)\nonumber \\ \leq c_1+n_1+c_3(m_0+n_1)+c_2(m_0+m_1+n_1) \label{B9}\\ 
c_2^2+c_3+c_2(1+c_1+c_3)+c_1(c_3+m_0)+(m_0+m_1)(m_1+n_1)\nonumber \\ \leq c_1^2+m_0+2c_2m_0+c_3m_0+m_1+c_1m_1+c_2m_1+(c_2+c_3)n_1\label{B10}\\
(c_2+c_3)(c_3-m_0)+2c_2n_1 \leq n_1^2 \label{B11}\\ 
c_2(c_2+c_3)+(2c_3+m_0)m_1 \leq m_0+c_3m_0+m_1^2+c_1(-1+c_2+m_1)\label{B12}\\ 
c_2^2+c_3+c_2(1+c_3)+m_0^2+c_1n_1+m_1(c_3+m_0+n_1) \leq (1+c_1+c_2+c_3)m_0+m_1+c_3n_1+c_2(m_1+n_1)\label{B13}\\ 
c_3^2+c_2(2+c_3-2m_0-m_1)+m_0(m_0+m_1)+(2m_0+m_1)n_1 \leq (2+c_3)m_0+n_1+c_1(-1+m_1+n_1)\label{B14}\\
c_3^2+c_2(1+c_3-2m_0+n_1)+m_0(m_0+n_1) \leq m_0+c_3 m_0+n_1^2\label{B15}\\
c_1+c_2+c_2^2+(c_1+c_2)c_3+m_0(m_0+2 n_1) \leq c_1^2+2(1+c_2)m_0+(c_1+c_2)n_1 \label{B16}
 \end{eqnarray}
\end{widetext}

%
%


\begin{thebibliography}{99}
\bibitem{foot1} The device independent success probability of Cabello's nonlocality argument (CNA) in quantum mechanics follows from \cite{Kunkri1}. Applying the same technique as used in \cite{Rabelo} to obtain the device independent success probability of Hardy's nonlocality argument, that of CNA is obtain.
\end{thebibliography}
\end{document}